\documentclass[10pt,letterpaper]{article}
\usepackage{opex3}

\usepackage{graphicx}%
\usepackage{dcolumn}
\usepackage{color}
\usepackage{amsmath}
\usepackage{here}

\begin{document}
\begin{sloppy}

\newcommand{\be}{\begin{equation}}
\newcommand{\ee}{\end{equation}}
\newcommand{\bea}{\begin{eqnarray}}
\newcommand{\eea}{\end{eqnarray}}%\newcommand{\rojo}[1]{\textcolor{red}{#1}}
\newcommand\bibtranslation[1]{English translation: {#1}}
\newcommand\bibfollowup[1]{{#1}}

\newcommand\pictc[5]{\begin{figure}
                       \centerline{
                       \includegraphics[width=#1\columnwidth]{#3}}
                  \protect\caption{\protect\label{fig:#4} #5}
                    \end{figure}            }
\newcommand\pict[4][1.]{\pictc{#1}{!tb}{#2}{#3}{#4}}
\newcommand\rpict[1]{\ref{fig:#1}}

\newcommand\leqt[1]{\protect\label{eq:#1}}
\newcommand\reqtn[1]{\ref{eq:#1}}
\newcommand\reqt[1]{(\reqtn{#1})}

\newcounter{Fig}
\newcommand\pictFig[1]{\pagebreak \centerline{
                   \includegraphics[width=\columnwidth]{#1}}
                   \vspace*{2cm}
                   \centerline{Fig. \protect\addtocounter{Fig}{1}\theFig.}}

\title{Tunable split-ring resonators for nonlinear negative-index metamaterials}

\author{Ilya V. Shadrivov, Steven K. Morrison, and Yuri S. Kivshar}

\address{Nonlinear Physics Centre, Research School of
Physical Sciences and Engineering \\
Australian National University,
Canberra ACT 0200, Australia}

\email{ivs124@rsphysse.anu.edu.au}

\homepage{http://wwwrsphysse.anu.edu.au/nonlinear}

\begin{abstract}
We study experimentally the dynamic tunability and self-induced
nonlinearity of split-ring resonators incorporating variable
capacitance diodes. We demonstrate that the eigenfrequencies of the
resonators can be tuned over a wide frequency range, and
significantly, we show that the self-induced nonlinear effects
observed in the varactor-loaded split-ring resonator structures can
appear at relatively low power levels.
\end{abstract}


\section{Introduction}

Specially engineered microstructured materials (or metamaterials)
that demonstrate many intriguing properties for the propagation of
electromagnetic waves such as negative refraction have been
discussed widely during recent years (see, e.g., the review
paper~\cite{OPN_costas} and references therein). Such materials have
been studied theoretically (see, e.g.,
Refs.~\cite{Pendry:1996-4773:PRL,Pendry:1999-2075:ITMT,Markos:2002-36622:PRE,Markos:2002-33401:PRB}
to cite a few) and also fabricated experimentally (see, e.g.,
Refs.~\cite{Smith:2000-4184:PRL,Bayindir:2002-120:APL,Parazzoli:2003-107401:PRL}).
The simplest composite material of this kind is created by a
three-dimensional lattice of metallic wires and split-ring
resonators (SRRs) with its unique properties associated with the
negative real parts of the magnetic permeability and dielectric
permittivity. Since the first theoretical paper by
Veselago~\cite{Veselago:1967-517:UFN}, such negative-index
metamaterials have also been described as {\em left-handed
materials}.

Within the microwave and millimeter frequency ranges, the composite
metamaterials created by SRR systems have been shown to possess
macroscopic negative-index (left-handed) properties and thus exhibit
peculiarities not found in natural materials. Recent strong
experimental efforts have been directed towards the attainment of
metamaterials with negative response in the
Terahertz~\cite{Linden:2004-1351:SCI} and even
optical~\cite{OPN_costas,Shalaev:2005-3356:OL} frequency domains.

Since the first extensive studies on negative-index metamaterials,
the attention of most researchers has been focused on the passive
control and linear properties of these composite structures, where
the effective parameters of the structure do not depend on the
intensity of the applied field or propagating electromagnetic waves.
However, to achieve the full potential of the unique properties of
the metamaterials requires the ability to dynamically control the
material's properties in real time through either {\em direct
external tuning} or {\em nonlinear responses}.

Dynamic control over metamaterials is a nontrivial issue since such
materials possess left-handed properties only in some finite
frequency range, which is basically determined by the geometry of
the structure. The possibility to control the effective parameters
of metamaterials using nonlinearity has recently been suggested in
Refs.~\cite{Zharov:2003-37401:PRL,Lapine:2003-65601:PRE} and
developed extensively in
Refs.~\cite{Shadrivov:2004-16617:PRE,Agranovich:2004-165112:PRB,
Lapine:2004-66601:PRE,Zharov:2005-91104:APL,Zharova:2005-1291:OE,
Shadrivov:2005-RS3S90:RS,Shadrivov:2006-529:JOSB,Gorkunov:2006-71912:APL}
where many interesting nonlinear metamaterials effects have been
predicted theoretically. The main reason for the expectation of
strong nonlinear effects in metamaterials is that the microscopic
electric field in the vicinity of the metallic particles forming the
left-handed structure can be much higher than the macroscopic
electric field carried by the propagating wave. This provides a
dynamic yet important physical mechanism for dramatic enhancing
nonlinear effects in left-handed materials. Moreover, changing the
intensity of the electromagnetic wave not only changes the material
parameters, but also allows switching between transparent
left-handed states and opaque dielectric states.

\pict[0.7]{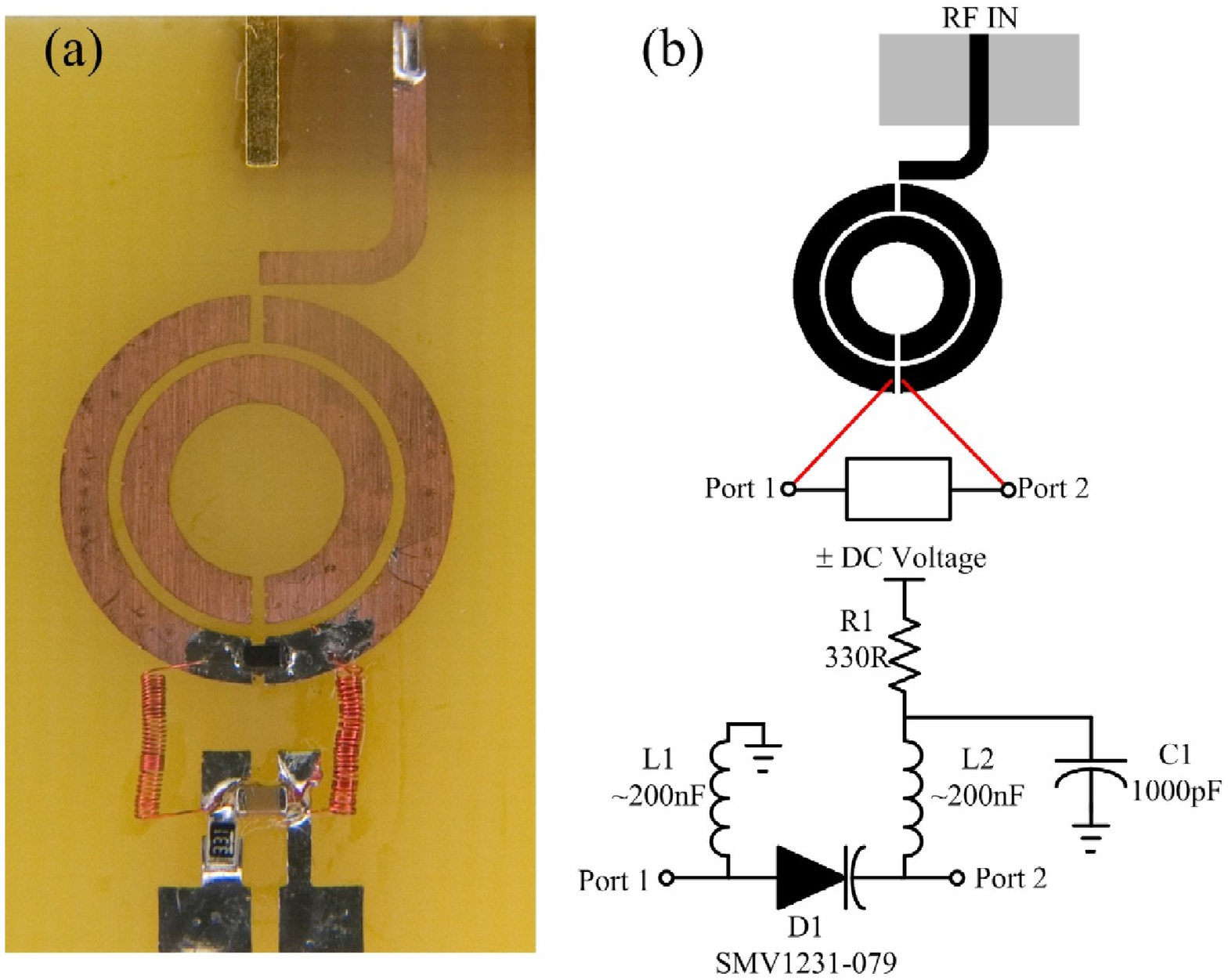}{geom_tuning}{Experimental structure used to
study tunability and nonlinearity of a varactor-loaded split-ring
resonator system. (a) Photography showing the fabricated SRR and
biasing circuitry used for direct voltage tuning of the resonant
frequency. (b) Top: electromagnetic representation of the SRR-diode
structure; bottom: schematic of the biasing circuit.}

In this paper, we make the first step toward the creation and study
of fully controlled, tunable nonlinear metamaterial systems through
the study of the tunability and self-induced nonlinear response of a
single SRR. The analysis of SRRs as tunable components within
extended composite systems is an extremely important step, and this
approach has, for example, been employed to study the effect of
disorder in SRR structures~\cite{Aydin:2004-5896:OE}. To achieve our
goals, we add a variable-capacitance diode (varicap) to the SRR
resonant structure. Tunability of the diode capacitance is achieved
by varying the width of its specifically doped P-N junction, and
associated depletion layer, through the application of a DC bias
voltage, or more attractively, through the self-action of the diode
with increased applied electromagnetic energy, giving rise to
dramatic self-induced nonlinear effects. Changes in the diode
capacitance alter the resonant conditions of the SRR producing
frequency shifts, which in turn adjust the effective magnetic
permeability of left-handed metamaterials. Using this methodology,
we experimentally establish a qualitative measure of the tunability
of an SRR, and confirm earlier predictions for the {\em dynamic
control} over the properties of SRRs by varying the intensity of the
incident electromagnetic signal.

\section{Tunable split-ring resonators}

Effective control over the resonant conditions of the SRR is
achieved by adding the capacitance of diodes in series with the
distributed capacitance of the outer ring of the SRR at a point of
maxima in the electric currents. This methodology, as well as the
motivation, is thus significantly different to previous works by Gil
\emph{et al.}~\cite{Gil:2004-1347:ELL,Gil:2006-2665:IEEE} that used
a varicap to obtain edge coupling {\em between the two rings} of a
uniquely designed SRR structure employed to form a microwave notch
filter. Indeed, the series application of the diode provides a
simple mechanism for both {\em tunability} and {\em nonlinearity}
suited to the formation of left-handed metamaterials, particularly
with the recent developments in magnetic thin-film and microwave
nonlinear materials. The symmetry and simplicity of our system also
lends itself to greater integration, allowing the structure to be
translated more readily to the THz and optical frequency domains.

To study the potential tunability and nonlinearity of metamaterials
we use a single archetype SRR, constructed on fibreglass (FR4,
$\epsilon_r \approx 4.4$) with copper metallization as shown in
Fig.~\rpict{geom_tuning}(a).  An additional gap is created in the
outer SRR ring for mounting of the Skywork's SMV1231-0790 varicap
diode. A DC voltage is used to achieve direct tuning of the diode.
The parasitic effects of the voltage supply are isolated from the
electromagnetic structure of the SRR via a biasing circuit. The
simple biasing circuit consist of wire-wound shunt inductors (L1,
L2) and de-coupling capacitor (C1), as illustrated in the schematic
of Fig.~\rpict{geom_tuning}(b). In the forward biasing regime a
series resistor (R1) is used to limit the forward bias current
through the diode and prevent damage; in the reverse bias region
this resistor has negligible effect on the circuit's operation.
Electromagnetic coupling to the SRR is accomplished with a short
trace probe near the SRR that is de-tuned to have a mostly flat
response in the SRR's operational bandwidth, allowing the SRR's
properties to dominate.  The SRR is energized and measured using a
20GHz Rohde and Schwartz vector network analyzer (ZVB20), wherein
resonant conditions are observed through a reduction in the
reflection coefficient ($S_{11}$) at resonance.
\pict{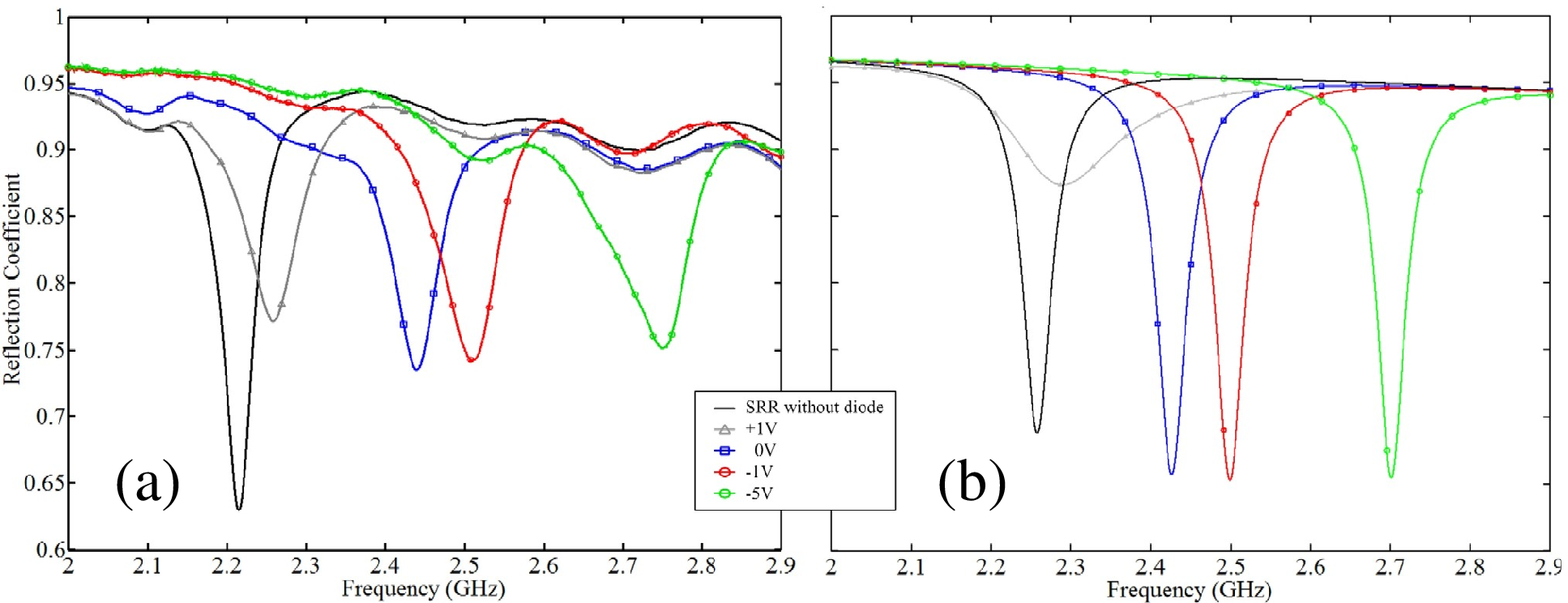}{tuning}{(a) Experimental and (b) numerical results
for the modulus of the reflection coefficient ($S_{11}$) as a
function of frequency for the SRR without the diode circuit, and
with the diode circuit under both positive and negative biasing
conditions.}

Two reference frequency responses are taken for the SRR: without the
diode and gap, and with the diode structure at zero bias voltage, as
shown in Fig.~\rpict{tuning} (a). An initial resonance located at
2.22GHz for the SRR is shifted to 2.44GHz with the diode structure.
This increase in resonant frequency is a result of the zero-bias
junction capacitance ($C_{J0}$) caused by the intrinsic depletion
layer of the diode. As the diode's capacitance is added in series to
the distributed capacitance of the SRR, it causes the total
capacitance to decrease and the resonance frequency to increase.
Application of a negative biasing voltage to the diode results in an
increase of the depletion layer and a subsequent decrease in the
diode's capacitance pushing the resonant frequency higher.
Conversely, a positive biasing voltage increases the diode's
capacitance and decreases the SRR's resonant frequency. However, due
to the current-voltage relationship of the diode it can only be
forward biased to a maximum voltage of 1.2 volts.

For a negative bias voltage of -10 volts (not shown) the resonant
frequency can be shifted to 2.9GHz, whereas for a positive voltage
of 1 volt the resonant frequency decreases to 2.27GHz. For this
particular varicap diode and SRR structure there is a tuning
bandwidth of 0.63GHz, equivalent to a tuning range of approximately
26\%.

We note here that this tuning range has not been optimized. Further
increases to this range could be achieved through improved materials
and impurity doping of the diode, bringing about an increase
capacitance ratio, in addition to greater integration of the diode
with the SRR structure. Indeed, Fig.~\rpict{tuning} (a) shows the
quality factor of the SRR's resonance is decreased with the
inclusion of the diode structure, resulting from the inherent series
resistance of the diode, and packaging and mounting parasitics; all
of which can be reduced through greater integration and more
advanced fabrication techniques.

\section{Numerical simulations}

To determine the optimal placement of the varicap diode, we employ
numerical simulations. We use a finite element method based
electromagnetic simulator in conjunction with a detailed SPICE model
of the varactor diode and bias circuit. S-parameters were extracted
from the SPICE model and included within the S-parameter
calculations of the SRR as a generic two-port device. Results of our
numerical simulations generated in this manner are presented in
Fig.~\rpict{tuning}(b) for several biased and unbiased conditions.

By comparing these results with the experimental curves, we observe
a good qualitative agreement between the measured and simulated
resonant frequency locations. However, the numerical results do not
reflect the voltage and frequency dependent changes in the quality
factors of the resonances. We believe that changes in the measured
quality factors of the resonances occur due to changes in the
parasitic series resistance of the diode and, to a lesser extent,
changes in the impedance mismatch of the diode and the SRR
structure. While the impedance mismatch effect is modeled
numerically, the variable series resistance is not, resulting in the
quality factor discrepancy. Greater accuracy of the simulated
results requires more knowledge of the diode's fabrication, which is
not available for the diode used in our experiment.

\pict[0.65]{fig03}{nonlinear_shift}{Experimentally
measured modulus of the SRR reflection coefficient ($S_{11}$) as a
function of frequency and applied input power from the vector
network analyzer. Power levels at the diode are significantly lower
due to input coupling (see text).}

\section{Nonlinear split-ring resonators}

Tunability of the SRR via an applied voltage is a valuable
mechanism; however the integration of the biasing network and
delivery of voltage sources, albeit possible, requires careful
design to prevent these components affecting the composite
performance of a LHM. Furthermore, the ability to change the
properties of a LHM in response to the power of the applied signal
leads to important effects such as dynamic control of the material's
opaqueness and the switching between left- and right-handed material
responses. To study these self-induced nonlinear effects we zero
bias the diode and sweep the applied power from the vector network
analyzer from 0dBm to 17dBm. The power levels at the diode are,
however, substantially lower due to the significant losses at the
input coupling probe.

As previously described theoretically in
Ref.~\cite{Zharov:2003-37401:PRL}, the nonlinear properties of the
SRR embedded with the varicap diode gives rise to eigenfrequency
shifts as the intensity of the applied electromagnetic wave is
varied. The self-induced nonlinear response of the SRR-diode
structure is illustrated in Fig.~\rpict{power_sweeps}(a), where the
dependence of the reflection coefficient on the applied signal is
observed for the zero-biased condition. The demonstrated reflection
coefficient dependence provides switching between high and low
reflection states, corresponding to high and low currents in the SRR
and thus the magnetic momenta. Control over the nonlinear response
can be achieved with static biasing of the diode as illustrated in
Fig.~\rpict{power_sweeps}(b), where a fixed reverse bias voltage of
-7 volts has been applied. Negative biasing results in an enhanced
response whereas positive biasing causes a reduced response arising
from the increased series resistance and diminished quality factor.
As the nonlinear effects are observed near the linear SRR resonance,
the frequency range where the SRR is strongly nonlinear shifts as
the bias voltage changes.

\section{Conclusions}

We have demonstrated experimentally both the direct real-time
tunability and self-induced nonlinear response of a varactor-loaded
split-ring resonator structure; a key element for the creation of a
nonlinear, negative-index metamaterial. The structure we have used
is based on a typical SRR design, and employs a commercially
available varactor diode placed at a maximum of the SRR's current
and in series with its distributed capacitance. Through the
application of a positive and negative biasing voltage, we have
observed more than half a GHz frequency tuning range; a tuning range
equivalent to 26\% of the zero-biased frequency. Using a zero-bias
diode, with and without an external biasing circuit and voltage
supply, we have demonstrated the ability of this structure to switch
between high and low reflection states associated with a
power-dependent frequency shift. These results suggest that
negative-index metamaterials constructed with SRRs and varactor
diodes can exhibit large tunability and nonlinearity, enabling many
unique and novel metamaterial properties not available within
nature. We believe our results will encourage the future design of
nonlinear, negative-index metamaterials within the microwave and
optical frequency domains.

\pict{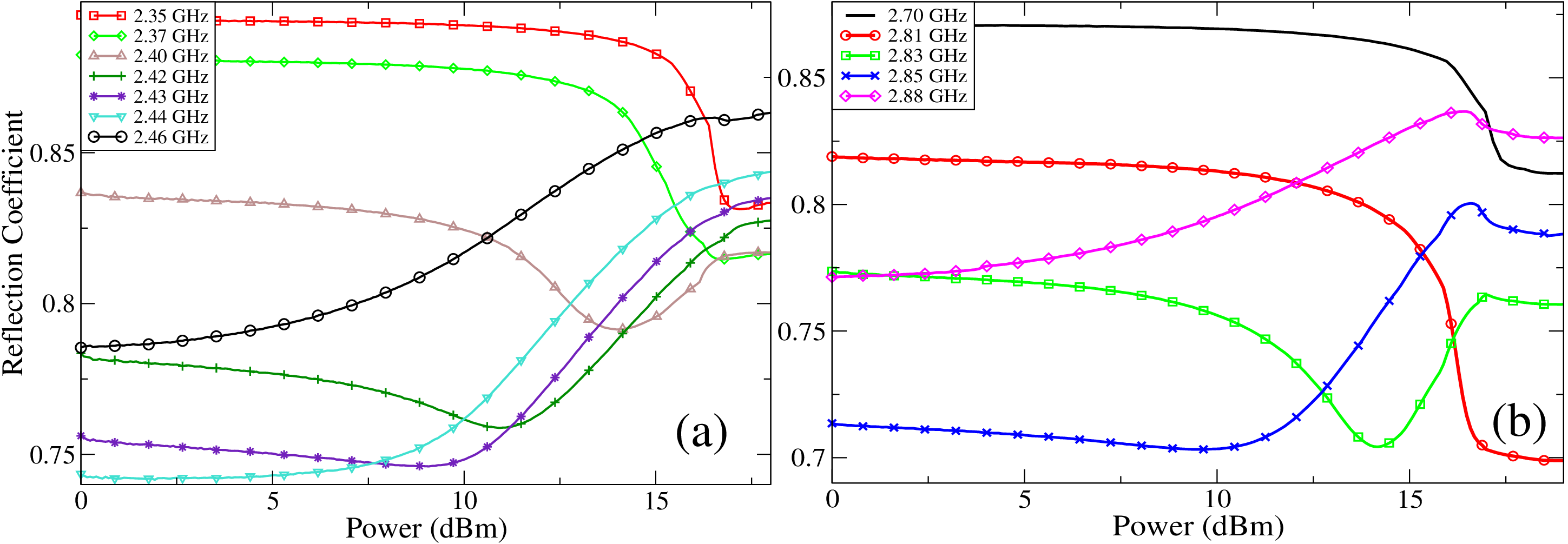}{power_sweeps}{Experimentally measured modulus of
the reflection coefficient ($S_{11}$) as a function of the output
power of vector network analyzer; (a) at several frequencies in the
vicinity of the linear resonance, and (b) zero bias conditions.
Power levels at the diode are significantly lower due to input
coupling (see details in the text).}

\section*{Acknowledgments}

This work has been supported by the Australian Research Council. We
thank C.M. Soukoulis, V. Shalaev, R. McPhedran, A. Mitchell, and M. Gorkunov for
useful and encouraging discussions.

\end{sloppy}

\begin{thebibliography}{10}

\bibitem{OPN_costas}  C. Soukoulis, ``Bending back light: The
science of negative index materials,'' Optics and Photonics News,
Vol. 17, No. 6, pp. 18-21 (2006), and references therein.

\bibitem{Pendry:1996-4773:PRL}
J.~B. Pendry, A.~J. Holden, W.~J. Stewart, and I. Youngs, ``Extremely low
  frequency plasmons in metallic mesostructures,'' Phys. Rev. Lett. {\bf 76,}
  4773--4776 (1996).

\bibitem{Pendry:1999-2075:ITMT}
J.~B. Pendry, A.~J. Holden, D.~J. Robbins, and W.~J. Stewart, ``Magnetism from
  conductors and enhanced nonlinear phenomena,'' IEEE Trans. Microw. Theory
  Tech. {\bf 47,} 2075--2084 (1999).

\bibitem{Markos:2002-36622:PRE}
P. Markos and C.~M. Soukoulis, ``Numerical studies of left-handed materials and
  arrays of split ring resonators,'' Phys. Rev. E {\bf 65,} 036622--8 (2002).

\bibitem{Markos:2002-33401:PRB}
P. Markos and C.~M. Soukoulis, ``Transmission studies of left-handed
  materials,'' Phys. Rev. B {\bf 65,} 033401--4 (2002).

\bibitem{Smith:2000-4184:PRL}
D.~R. Smith, W.~J. Padilla, D.~C. Vier, S.~C. Nemat~Nasser, and S. Schultz,
  ``Composite medium with simultaneously negative permeability and
  permittivity,'' Phys. Rev. Lett. {\bf 84,} 4184--4187 (2000).

\bibitem{Bayindir:2002-120:APL}
M. Bayindir, K. Aydin, E. Ozbay, P. Markos, and C.~M. Soukoulis, ``Transmission
  properties of composite metamaterials in free space,'' Appl. Phys. Lett. {\bf
  81,} 120--122 (2002).

\bibitem{Parazzoli:2003-107401:PRL}
C.~G. Parazzoli, R.~B. Greegor, K. Li, B.~E.~C. Koltenbah, and M. Tanielian,
  ``Experimental verification and simulation of negative index of refraction
  using Snell's law,'' Phys. Rev. Lett. {\bf 90,} 107401--4 (2003).

\bibitem{Veselago:1967-517:UFN}
V.~G. Veselago, ``The electrodynamics of substances with simultaneously
  negative values of epsilon and mu,'' Usp. Fiz. Nauk {\bf 92,} 517--526 (1967)
  (in Russian) [English translation: Phys. Usp. {\bf 10,} 509--\& (1968)].

\bibitem{Linden:2004-1351:SCI}
S. Linden, C. Enkrich, M. Wegener, J. Zhou, T. Koschny, and C.~M. Soukoulis,
  ``Magnetic response of metamaterials at 100-terahertz,'' Science {\bf 306,}
  1351--1353 (2004).

\bibitem{Shalaev:2005-3356:OL}
V.~M. Shalaev, W. Cai, U.~K. Chettiar, H.~K. Yuan, A.~K. Sarychev, V.~P.
  Drachev, and A.~V. Kildishev, ``Negative index of refraction in optical
  metamaterials,'' Opt. Lett. {\bf 30,} 3356--3358 (2005).

\bibitem{Zharov:2003-37401:PRL}
A.~A. Zharov, I.~V. Shadrivov, and Yu.~S. Kivshar, ``Nonlinear properties of
  left-handed metamaterials,'' Phys. Rev. Lett. {\bf 91,} 037401--4 (2003).

\bibitem{Lapine:2003-65601:PRE}
M. Lapine, M. Gorkunov, and K.~H. Ringhofer, ``Nonlinearity of a metamaterial
  arising from diode insertions into resonant conductive elements,'' Phys. Rev.
  E {\bf 67,} 065601--4 (2003).

\bibitem{Shadrivov:2004-16617:PRE}
I.~V. Shadrivov, A.~A. Sukhorukov, Yu.~S. Kivshar, A.~A. Zharov, A.~D.
  Boardman, and P. Egan, ``Nonlinear surface waves in left-handed materials,''
  Phys. Rev. E {\bf 69,} 16617--9 (2004).

\bibitem{Agranovich:2004-165112:PRB}
V.~M. Agranovich, Y.~R. Shen, R.~H. Baughman, and A.~A. Zakhidov, ``Linear and
  nonlinear wave propagation in negative refraction metamaterials,'' Phys. Rev.
  B {\bf 69,} 165112--165117 (2004).

\bibitem{Lapine:2004-66601:PRE}
M. Lapine and M. Gorkunov, ``Three-wave coupling of microwaves in metamaterial
  with nonlinear resonant conductive elements,'' Phys. Rev. E {\bf 70,}
  66601--8 (2004).

\bibitem{Zharov:2005-91104:APL}
A.~A. Zharov, N.~A. Zharova, I.~V. Shadrivov, and Yu.~S. Kivshar,
  ``Subwavelength imaging with opaque nonlinear left-handed lenses,'' Appl.
  Phys. Lett. {\bf 87,} 091104--3 (2005).

\bibitem{Zharova:2005-1291:OE}
N.~A. Zharova, I.~V. Shadrivov, A.~A. Zharov, and Yu.~S. Kivshar, ``Nonlinear
  transmission and spatiotemporal solitons in metamaterials with negative
  refraction,'' Optics Express {\bf 13,} 1291--1298 (2005).

\bibitem{Shadrivov:2005-RS3S90:RS}
I.~V. Shadrivov, A.~A. Zharov, N.~A. Zharov, and Yu.~S. Kivshar, ``Nonlinear
  left-handed metamaterials,'' Radio Science {\bf 40,} RS3S90 (2005).

\bibitem{Shadrivov:2006-529:JOSB}
I.~V. Shadrivov, A.~A. Zharov, and Yu.~S. Kivshar, ``Second-harmonic generation
  in nonlinear left-handed metamaterials,'' Journal of the Optical Society of
  America B (Optical Physics) {\bf 23,} 529--534 (2006).

\bibitem{Gorkunov:2006-71912:APL}
M.~V. Gorkunov, I.~V. Shadrivov, and Yu.~S. Kivshar, ``Enhanced parametric
  processes in binary metamaterials,'' Appl. Phys. Lett. {\bf 88,} 71912--3
  (2006).

\bibitem{Aydin:2004-5896:OE} K. Aydin, K. Guven, N. Katsarakis, C.M.
Soukoulis, and E. Ozbay, ``Effect of disorder on magnetic resonance
band gap in split-ring resoanntor structures,'', Opt. Exp. {\bf 12},
5896-5901 (2004).

\bibitem{Gil:2004-1347:ELL}
I. Gil, J. Garcia~Garcia, J. Bonache, F. Martin, M. Sorolla, and R. Marques,
  ``Varactor-loaded split ring resonators for tunable notch filters at
  microwave frequencies,'' Electron. Lett. {\bf 40,} 1347--1348 (2004).

\bibitem{Gil:2006-2665:IEEE} I. Gil, J. Bonache, J. Garc\'ia-Garc\'ia, F. Mart\'in, ``Tunable
metameterial transmission lines based on varactor-loaded split-ring
resonantors,'' IEEE Trans. on Microwave Theory and Techn., {\bf 54},
2665 - 2674 (2004).

\end{thebibliography}
\end{document}